# Quantum Enhanced Microrheology of a Living Cell

Michael A. Taylor*, Jiri Janousek**, Vincent Daria**, Joachim Knittel*, Boris Hage**, Hans-A. Bachor**, and Warwick P. Bowen*

\* Centre for Engineered Quantum Systems, University of Queensland, St Lucia, Queensland 4072, Australia
\** Department of Quantum Science, Australian National University, Canberra, ACT 0200, Australia

**Abstract**

We demonstrate the first biological measurement with precision surpassing the quantum noise limit. Lipid particles within a living yeast cell are tracked with sub-shot noise sensitivity, thereby revealing the biological dynamics of the cellular cytoplasm.

## I. Introduction

Quantum metrology allows high sensitivity measurements to proceed with a lower light intensity than classically possible [1]. A particularly important frontier for this technology is in biological measurements, where photochemical interactions often disturb biological processes and can damage the specimen [2]. Here we report the first demonstration of biological measurement with precision surpassing the quantum noise limit [3]. We used amplitude squeezed light to perform microrheology experiments within *Saccharomyces cerevisiae* yeast cells with precision surpassing the quantum noise limit by 42%. This entailed tracking the thermal motion of naturally occurring lipid granules, which is determined by the mechanical properties of the surrounding cytoplasm and the embedded polymer networks. From this motion the viscoelastic moduli of the surrounding cytoplasm could be determined in real time, with squeezed light allowing a 64% higher measurement rate than possible classically, improving the temporal resolution of the biological dynamics of the cellular cytoplasm [5]. The approach presented here is widely applicable, extending the reach of quantum enhanced measurement to many dynamic biological processes. Furthermore, by demonstrating that biological measurements can be improved using quantum correlated light, our results pave the way to a broad range of applications in areas such as two-photon microscopy, super-resolution, and absorption imaging [1].

## II. Particle tracking method

To enable the reported results, we developed a new laser based microscopy system which extended previous methods used to track the motion of highly reflective mirrors with non-classical light to measurements of microscopic particles with non-paraxial fields (shown in Fig. 1). This is an optical tweezers setup with a number of significant modifications. Firstly, dark-field illumination is used [6], which intrinsically causes only side-scattered light to enter the measurement. Because large objects predominantly scatter light forward, this feature ensures that when measuring motion in yeast cells, most of the captured light has scattered from a lipid granule. Secondly, this illumination is stroboscopically pulsed, allowing an optical lock-in measurement which made biological dynamics in the critical Hz-kHz frequency range accessible. This straightforward technique allowed quantum enhancement over a frequency range which reached as low as the range reported for squeezed light sources developed for gravity wave interferometers [4], without changing the quantum limit on sensitivity [7]. Thirdly, a self-homodyne measurement was performed, with the position extracted from interference between a shaped local oscillator field and scattered light rather than by using the conventional quadrant detector. This allowed the local oscillator field shape to be optimized independently of the probe or trapping fields, and when amplitude squeezed, any spatial mode perturbations occurring during optical propagation were applied equally to both the squeezing and local oscillator, ensuring perfect overlap at detection.

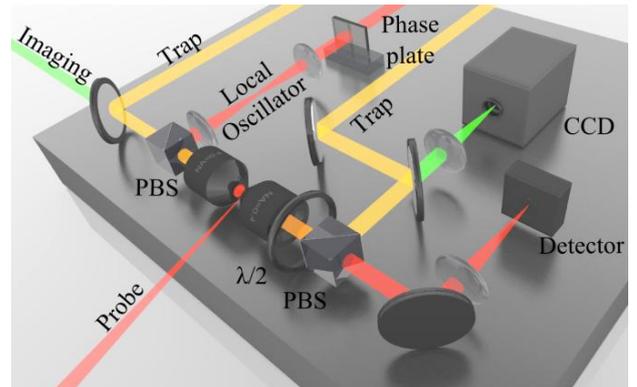

Fig. 1. Layout of the particle tracking experiment. Counter-propagating trap fields (yellow) confined the particles, and are isolated from the detection with polarizing beamsplitters (PBS). A separate probe field illuminates the particles, and scattered light from this mixes with a local oscillator to reveal particle position. A separate green field is also used to image the particles onto a CCD camera.

With this apparatus, we tracked 2 μm diameter silica beads in water, and confirmed that we could measure Brownian motion with the lock-in technique (Fig. 2 a, b). With squeezed light, we could enhance the sensitivity by up to 2.7 dB (Fig. 2 c).

This work was supported by the Australian Research Council Discovery Project Contract No. DP0985078.

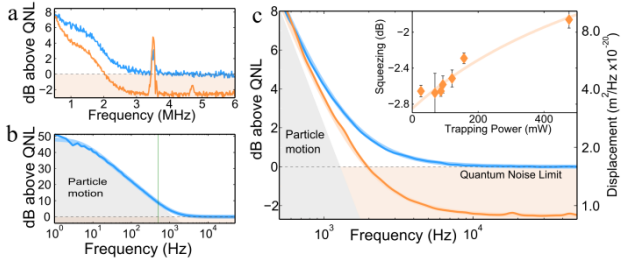
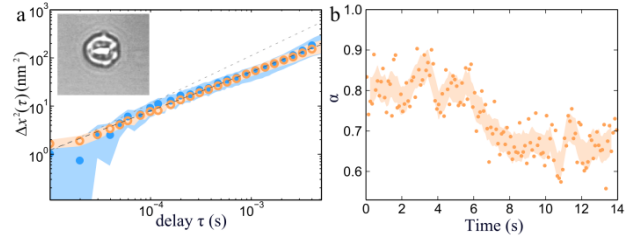

Fig. 2. The experimental results are shown here. (a) The lock-in technique allowed quantum noise limited measurements despite the presence of low frequency noise. (b) Silica beads in water were found to follow the expected Brownian motion. (c) When using squeezed light, the beads could be tracked with 2.7 dB improved precision. The inset shows the degradation in squeezing due to increasing trap power. Here, the blue lines are recorded with coherent light, and the orange lines with squeezed light.

Then, lipid granules within yeast cells were tracked, with a quantum enhancement of 2.4 dB achieved. There, the motion was analyzed in terms of the mean squared displacement, which is defined in terms of the position $x$ at time $t$ as

$$\Delta x^2(\tau) = \langle (x(t) - x(t-\tau))^2 \rangle = 2D\tau^\alpha, \quad (1)$$

where $\tau$ is a delay between measurement, $D$ is the diffusion constant, and the parameter $\alpha$ determines the diffusive regime. Brownian motion follows when $\alpha=1$, and when $\alpha<1$, the motion is termed subdiffusive. Subdiffusion occurs when the surrounding cytoplasm exhibits both viscosity and elasticity, with the value of $\alpha$ determined by the ratio of viscosity to elasticity [8]. The viscoelasticity of the cytoplasm is extremely significant property, as it affects the rate at which reactants can combine within the cell [9].

Our measurements of lipid motion were fitted to Eq. (1) to determine $\alpha$, as shown in Fig. 3 a. This showed that the lipid motion was consistently subdiffusive, as previous experiments have found [10], and that $\alpha$ changed dynamically [5] as the lipid granules moved within the cytoplasm (Fig. 3 b). Since the cytoplasm is inhomogeneous, the lipid particle interacts with a different environment when it moves, and the local viscoelasticity can vary spatially by over an order of magnitude [11]. Squeezed light improved the precision with which $\alpha$ could be determined by 22%, or equivalently, allowed the measurement rate to increase by 64%. As such, squeezed light improved our ability to resolve dynamic changes of the viscoelastic properties of the cell.

## III. Conclusion

We have demonstrated a novel microscopy technique which is capable of breaking the quantum noise limit in measurements. This was applied in measurements of the mechanical properties of a living cell, thereby successfully demonstrating the first biological measurement to break the quantum limit.

Fig. 3. (a) Lipid granules within yeast cells were tracked, and their mean squared displacement analyzed, from which the diffusive parameter α could be determined. (b) A series of measurements showed that α dynamically changes on sub-second timescales, and squeezed light could allow the measurement rate to be increased by 64%.

This work was supported by the Australian Research Council Discovery Project Contract No. DP0985078.